\documentclass[a4paper, 12pt]{article}
\usepackage{graphicx}
\usepackage{amsmath}
\usepackage{amssymb}
\usepackage{subfigure}

\def \be {\begin{equation}}
\def \ee {\end{equation}}
\def \bea {\begin{eqnarray}}  
\def \eea {\end{eqnarray}} 
\def \mea {\nonumber\\}

\begin{document}    
\begin{titlepage}

\title{Planck's Other Quanta: Corpuscular Electrodiffusion}     
     
\author{
L. Bass\footnote{{\em Email:} lb@maths.uq.edu.au}\,\,\,and A.J. Bracken\footnote{{\em Email:} a.bracken@uq.edu.au} 
\\Department of Mathematics\\    
The University of Queensland\\Brisbane 4072, Australia}

\date{}     
\maketitle     
\begin{abstract}
When B\"acklund transformations are applied repeatedly to any solution of the nonlinear equations describing 
electrodiffusion through a liquid junction separating two infinite well-stirred layers, 
they give rise to quantized ionic fluxes across 
the junction.  A particular exact solution is shown to imply that these fluxes consist of quanta of electric charge.  
\end{abstract}

     
\end{titlepage}

\section{Introduction}
Since the pioneering works of Nernst \cite{nernst} and Planck \cite{planck}, transport of charged ions across liquid junctions 
has played a fundamental role in a variety of natural systems.
An extension  of the  
Nernst-Planck model by one of us \cite{bass1,bass2} 
incorporated the effect of the
electric field that develops within a junction in response to diffusional separation of
ions carrying different charges.  The resulting nonlinearity is essential to what follows. 

In its simplest form, the model
deals with steady one-dimensional transport of two ionic species with equal and opposite charges,
across an infinite slab occupying $0\leq x \leq \delta$.  
The governing system of differential equations is
\bea
c_+\,'(x)= ({z}e/k T)\,E(x)\,c_+(x)-\Phi_+/D_+\,,
\mea\mea
c_-\,'(x)= -({z}e/k T)\,E(x)\,c_-(x)-\Phi_-/D_-\,,
\mea\mea
E\,'(x)=(4\pi {z} e/\epsilon)\left[c_+(x)-c_-(x)\right]\qquad
\label{system1}
\eea
for $0<x<\delta$.  Here $c_{\pm}(x)$ denote the concentrations of the two ionic species, $\Phi_{\pm}$ their steady 
(constant) fluxes in the $+x$-direction, $D_{\pm}$ their diffusion coefficients, and $z$ their common valence, 
while $E(x)$ denotes the induced electric field,
$k$ Boltzmann's constant, $e$
the electronic charge, and $T$
the ambient absolute temperature within the solution in the slab.  
An important auxiliary quantity is the 
electric current-density
\bea
J=J_++J_-\,,\quad J_{\pm}=\pm {z}e\,\Phi_{\pm}\,.
\label{current1}
\eea
Elimination of $c_+(x)$ and $c_-(x)$ from \eqref{system1} results in a second-order nonlinear equation for $E(x)$ that
can be transformed into a Painlev\'e equation of the second kind \cite{ince}  that has been studied extensively in this context \cite{rogers1,rogers6}.  
Here we prefer to keep to the set \eqref{system1} and focus on the behaviour of all the physical variables.  
   
A feature \cite{rogers1,rogers6,bracken1} of the system \eqref{system1} is that it admits an auto-B\"acklund 
transformation \cite{rogers7} ${\cal B}$ 
and its inverse ${\cal B}^{-1}$.  Thus, given any one solution
\bea
{\cal S}^{(0)}=(c_+^{(0)}(x),\,c_-^{(0)}(x),\,E^{(0)}(x),\,\Phi_+^{(0)},\,\Phi_-^{(0)})\,,
\label{seed}
\eea
a second solution is given by 
\bea
{\cal S}^{(1)}={\cal B}({\cal S}^{(0)})=({c}_+^{(1)}(x),\,{c}_-^{(1)}(x),\,{E}^{(1)}(x),\,{\Phi}_+^{(1)},\,{\Phi}_-^{(1)})
\label{B_soln1}
\eea
where
\bea
{c}_+^{(1)}(x)=c_-^{(0)}(x)-\frac{\epsilon}{2\pi z e D_+}\,\frac{\Phi_+^{(0)}\,E^{(0)}(x)}{c_+^{(0)}(x)}
+\frac{\epsilon k T}{2\pi z^2e^2}\,\left( \frac{\Phi_+^{(0)}}{D_+c_+^{(0)}(x)}\right)^2\,,
\mea\mea
c_-^{(1)}(x)=c_+^{(0)}(x)\,,\quad E^{(1)}(x)=-E^{(0)}(x)+\frac{2kT}{zeD_+}\,\frac{\Phi_+^{(0)}}{c_+^{(0)}(x)}\,,
\mea\mea
\Phi_+^{(1)}=2\Phi_+^{(0)}+(D_+/D_-)\Phi_-^{(0)}\,,\quad \Phi_-^{(1)}=-(D_-/D_+)\Phi_+^{(0)}\,,
\label{B_soln2}
\eea
and a third solution is given by
\bea
{\cal S}^{(-1)}={\cal B}^{-1}({\cal S}^{(0)})=({c}_+^{(-1)}(x),\,{c}_-^{(-1)}(x),\,{E}^{(-1)}(x),\,{\Phi}_+^{(-1)},\,{\Phi}_-^{(-1)})\,,
\label{Binv_soln1}
\eea
where
\bea
\mea
{c}_-^{(-1)}(x)=c_+^{(0)}(x)+\frac{\epsilon}{2\pi z e D_-}\,\frac{\Phi_-^{(0)}\,E^{(0)}(x)}{c_-^{(0)}(x)}
+\frac{\epsilon k T}{2\pi z^2e^2}\,\left( \frac{\Phi_-^{(0)}}{D_-c_-^{(0)}(x)}\right)^2\,,
\mea\mea
c_+^{(-1)}(x)=c_-^{(0)}(x)\,,\quad E^{(-1)}(x)=-E^{(0)}(x)-\frac{2kT}{zeD_-}\,\frac{\Phi_-^{(0)}}{c_-^{(0)}(x)}\,,
\mea\mea
\Phi_-^{(-1)}=2\Phi_-^{(0)}+(D_-/D_+)\Phi_+^{(0)}\,,\quad \Phi_+^{(-1)}=-(D_+/D_-)\Phi_-^{(0)}\,.
\label{Binv_soln2}
\eea
Repeated application of ${\cal B}$ and ${\cal B}^{-1}$ on any given {\em seed solution} 
${\cal S}^{(0)}$ generates a sequence of solutions
\bea
{\cal S}^{(n)}\,,\quad n=0,\,\pm1,\,\pm2,\,\dots
\label{sequence}
\eea
that is doubly infinite in general.  

%

Considering repeated transformations as in \eqref{B_soln2}, \eqref{Binv_soln2} and \eqref{sequence}, we deduce that 
\bea
\Phi_+^{(n)} =(n+1)\Phi_+^{(0)} + n(D_+/D_-)\Phi_-^{(0)}\,,
\mea\mea
\Phi_-^{(n)}=-(n-1)\Phi_-^{(0)} -n (D_-/D_+)\Phi_+^{(0)}\,,
\label{fluxes1}
\eea
so that 
\bea
J_+^{(n)} =(n+1)J_+^{(0)} - n(D_+/D_-)J_-^{(0)}\,,
\mea\mea
J_-^{(n)}=-(n-1)J_-^{(0)} +n (D_-/D_+)J_+^{(0)}\,,
\label{current2}
\eea
and hence
\bea 
J^{(n)}=J^{(0)}+ n\Delta J\,,\quad 
\Delta J={\tilde z}e(D_++D_-)\left\{ \frac{\Phi_+^{(0)}}{D_+}+\frac{\Phi_-^{(0)}}{D_-}\right\}\,,
\label{current3}
\eea
where 
\bea
J^{(0)}=J_+^{(0)}+J_-^{(0)}=ze\left(\Phi_+^{(0)}-\Phi_-^{(0)}\right)\,.
\label{current4}
\eea
This mathematical phenomenon has been called `B\"acklund flux-quantization' \cite{bracken1}.  
It is associated with any 
solution ${\cal S}^{(0)}$ of \eqref{system1} taken as seed, in particular any exact solution, 
whether known  \cite{rogers6} or as yet unknown.  
Physically the phenomenon  may be related, for example, to discrete fluxes such as occur in electrochemical 
action potentials across excitable membranes \cite{bass3}  in plants or animals.  
\section{Quantal structure underlying solutions}
A notable feature of \eqref{current3} is that increments of fluxes pertaining to successive values of $n$ 
are all equal and all determined at $n=0$, despite radical differences  between the solutions of \eqref{system1} associated with those fluxes at different $n$ values. What common element of such disparate solutions can engender consistency with equality of the flux increments? 
A special exact solution implies that the common element is the quantum of electric charge, as we now show.

Consider the exact solution ${\cal S}^{(0)}$ of \eqref{system1} with
\bea
c_+^{(0)}(x)=c_-^{(0)}(x)=c_0+(c_1-c_0)x/\delta\,,\quad E^{(0)}(x)=0\,,
\mea\mea
\Phi_+^{(0)}=D_+(c_0-c_1)/\delta\,,\quad \Phi_-^{(0)}=D_-(c_0-c_1)/\delta\,.
\label{planck3}
\eea 
Here $c_0>c_1>0$ are constants.   
This is a special case of a solution given by Planck \cite{planck} for the Nernst-Planck model, 
one that happens to provide also a solution of \eqref{system1}.  It describes a steady-state 
situation where there the electric field vanishes throughout the slab,
so that the ionic transport is purely diffusive, and charge neutrality is maintained throughout, including at the two faces. 
From this seed solution other exact solutions can be derived by B\"acklund transformations, as in \eqref{B_soln1}--\eqref{sequence}.  
To simplify the discussion, suppose that the diffusion coefficients are equal: $D_+=D_-=D$.  Then 
\bea
\Phi_+^{(0)}=\Phi_-^{(0)}= 
D(c_0-c_1)/\delta\,
\label{fluxes3}
\eea
and hence \eqref{current3} and \eqref{current4} give 
\bea
J_+^{(0)}=-J_-^{(0)}=zeD(c_0-c_1)/\delta\,,
\mea\mea
 J^{(0)}=0\,,\quad
\Delta J=4zeD(c_0-c_1)/\delta\,.
\label{planck4}
\eea

It follows from \eqref{fluxes3} that in a time $\tau$, the number of positive ions (equal to the number of negative ions)  
flowing in the $+x$-direction (down the concentration gradients) 
across an area $A$ perpendicular to the $x$-axis is given by 
\bea
n_+=n_-=\Phi_{\pm}^{(0)}A\tau=D(c_0-c_1)A\tau/\delta\,.
\label{numbers1}
\eea
Consider the situation where $n_+=n_-\approx 1$.  Each of these two ions performs a random walk across the junction in a time
\bea
\tau\approx \delta^2/2D\,.
\label{time1}
\eea
With this value for $\tau$,  setting $n_+=n_-\approx 1$ in \eqref{numbers1} means choosing $A$ so that 
\bea
A\approx 2/(c_0-c_1)\delta\,.
\label{area}
\eea
Then it follows from \eqref{current4} and \eqref{numbers1} that this single positively ({\em resp.} negatively) charged ion carries 
a charge $ze$ ({\em resp.} $-ze$) down the concentration gradients, and the nett charge across
$A$ is zero, consistent with the vanishing of the current density
$J^{(0)}$ as in \eqref{planck4}.  Note that for these values of $A$ and $\tau$,  
\bea
\Delta J A\tau=4 ze\,.
\label{DeltaJ}
\eea

Proceeding from ${\cal S}^{(0)}$ to solutions ${\cal S}^{(n)}$, $n=\pm 1,\,\pm 2,\,\dots$, we see from \eqref{current3} that $\Delta J$ is 
unchanged despite changes in concentrations, appearance of electric fields, and loss of electroneutrality within the junction 
as evident in \eqref{B_soln2} and 
\eqref{Binv_soln2}.   Thus electrical charges $Q^{(n)}$ transferred in time $\tau$ are, from \eqref{current3} and \eqref{DeltaJ},
\bea
Q^{(n)} =n\Delta J\,A\,\tau=4nze\,,\quad n=0,\,\pm 1,\,\pm 2,\,\dots\,,
\label{Qdef}
\eea
so that the quanta implied here are quadruples of $ze$.  [When $D_+\neq D_-$ the same incremental charges $Q^{(n)}$ 
are transferred across the area $A$   
in the time $\tau '= 2\tau_+\tau_- /(\tau_++\tau_-)$, where $\tau_{\pm}$ is defined as in \eqref{time1}, with $D_{\pm}$ replacing $D$.]

To elucidate this result in terms of transfers of positive and negative ions, 
consider the B\"acklund-transformed solution ${\cal S}^{(1)}$ as in \eqref{B_soln2}, with Planck's solution \eqref{planck3} 
as ${\cal S}^{(0)}$.  Keeping the same values for $A$ and $\tau$ we have from \eqref{current3}
\bea
J^{(1)}A\tau =\left(J^{(0)}+\Delta J\right)A\tau=4 ze\,.
\label{Bcharge1}
\eea
Furthermore, from \eqref{current2}, 
\bea
J_+^{(1)}A\tau=3ze\,,\quad J_-^{(1)}A\tau=ze\,,
\label{Bcharge2}
\eea
Thus ${\cal S}^{(1)}$ describes a situation in which $3$ positive ions are transported across $A$ 
in the $+x$-direction in time $\tau$,
while
$1$ negative ion is transported in the $-x$-direction.  Note that transport in this situation is no longer by diffusion alone, as the electric field
is now non-vanishing.   

The inverse-B\"acklund transformed solution ${\cal S}^{(-1)}$ describes the charge-conjugate situation, with
\bea
J^{(-1)}A\tau =-4 ze\,,\qquad\qquad
\mea\mea
J_+^{(-1)}A\tau=-3ze\,,\quad J_-^{(-1)}A\tau=-ze\,.
\label{Binvcharge2}
\eea
More generally, according to \eqref{current2}, \eqref{current3} and \eqref{DeltaJ}, 
the solution ${\cal S}^{(n)}$ describes a situation in which 
\bea
J^{(n)}A\tau=4nze\,,\qquad\qquad\qquad\quad
\mea\mea
J_+^{(n)}A \tau=(2n+1) ze\,,\quad J_-^{(n)}A\tau=(2n-1)ze\,,
\label{Bcharges3}
\eea
for $n=0,\,\pm 1,\,\pm 2,\,\dots$, corresponding to $2n+1$ positively charged ions transported across $A$ in the $+x$-direction, and 
$2n-1$ negatively charged ions in the $-x$-direction.  In this way, we see that B\"acklund flux-quantization directly reflects the quantization
of electric charge on the two ionic species.   
%

Consider ${\cal S}^{(1)}$ in more detail.  From \eqref{planck3} and \eqref{B_soln2}
we have
\bea
\frac{c_+^{(1)}(x)}{c_0}=1+\left(\frac{c_1}{c_0}-1\right)\frac{x}{\delta}+ \frac{\epsilon E^{(1)}(x)^2}{8\pi kT c_0}\,,
\label{excited1}
\eea
where
\bea
E^{(1)}(x)= \left(\frac{2kT}{ze\delta}\right)\,\frac{c_0-c_1}{c_0+(c_1-c_0)x/\delta}
\label{excited2}
\eea
with similar expressions for $c_-^{(-1)}(x)$ and $E^{(-1)}(x)$.  The third (dimensionless) term on the RHS of \eqref{excited1}
typically has a value $10^{-10}$ or less \cite{bass2}, yet such terms give rise to 
the B\"acklund-generated exact solutions ${\cal S}^{(1)}$ and ${\cal S}^{(-1)}$ with 
Planck's solution \eqref{planck3} as seed.   
While these derived solutions collapse onto the seed 
solution as $\epsilon E^{(\pm 1)}(x)^2/kTc_0$ goes to zero, the quantal structure \eqref{Bcharge1}, \eqref{Bcharge2}, \eqref{Binvcharge2} 
remains unaffected.  
%
%
%
%
%
%
%
\section{Historical note}
The theory of electrodiffusion was initiated by Planck \cite{planck} a decade before his 
quantization of black-body radiation \cite{planck2}.  
The extensive developments of the latter were not slowed by mathematical complexities.  In contrast, Planck's work on electrodiffusion
had first to be supplemented by essential nonlinearities \cite{bass1,bass2} before 
the pioneering 
ideas of B\"acklund \cite{backlund} could be  applied \cite{rogers1,rogers6,bracken1} to reveal 
at least a fraction of the wealth of exact solutions of \eqref{system1}, based only on the particular seed solution \eqref{planck3}, but 
sufficient to imply the quanta of electric charge, as in \eqref{Qdef} and \eqref{Bcharges3}.  Despite the passage of time, these surprising products of the nonlinearity of the continuum theory \eqref{system1} may be regarded as Planck's other quanta.


\end{document}